\documentclass[fleqn,twoside]{article}
\usepackage{espcrc2}

\usepackage{graphicx}
\usepackage[figuresright]{rotating}

\newcommand{\AmS}{{\protect\the\textfont2
  A\kern-.1667em\lower.5ex\hbox{M}\kern-.125emS}}

\title{Instabilities in Molecular Dynamics Integrators used in Hybrid Monte Carlo Simulations}

\author{B. Jo\'o {\em UKQCD Collaboration}}

\begin{document}

\begin{abstract}
We discuss an instability in the leapfrog integration algorithm,
widely used in current Hybrid Monte Carlo (HMC) simulations of lattice
QCD. We demonstrate the instability in the simple harmonic oscillator
(SHO) system where it is manifest. We demonstrate the instability in
HMC simulations of lattice QCD with dynamical Wilson-Clover
fermions and discuss implications for future simulations of lattice
QCD.
\vspace{1pc}
\end{abstract}

\maketitle

\section{INTRODUCTION}
The subject of instabilities in molecular dynamics (MD) integrators is
not new. Studies of the phenomenon for the case of lattice QCD
have been reported in \cite{Edwards,Joo}. In \cite{Edwards} the
lattices used were small ($4^4$ sites).  The onset of
instabilities was investigated as a function of the integration step
size ($\delta \tau$) for heavy and light quark masses.  For
both masses studied the instabilities set in at some critical step
size.  The critical value of $\delta \tau$ was smaller for the lighter
mass than for the heavier one. However, in both cases $\delta \tau$
was still so large that the energy change ($\delta H$) from the
step size induced errors along an MD trajectory was
sufficiently great to make simulations impractical in the unstable
region of parameter space anyway.

Our study \cite{Joo} using larger lattices and quark masses that are
relatively light (by current standards) encountered instabilities at
step sizes that are small enough to be relevant for large scale
simulations. Our motivation to summarise these results is to highlight
the impact these instabilities may have on future simulations using MD
integrators as components, in particular HMC.

This article is organised as follows. In section \ref{s:SHO} we demonstrate
the instability in the simple harmonic oscillator (SHO) system. At this point
we will restate the hypothesis of \cite{Edwards} to explain 
the onset of the instability. We present our main  results in section \ref{s:OurData} and show that they are completely consistent with this hypothesis.
Finally in section \ref{s:Conclusions} we will draw our conclusions.

\section{SHO RESULT AND HYPOTHESIS}\label{s:SHO}
In the case of the SHO system a leapfrog update
can be written as:
\begin{equation}
\scriptsize
\left[ \hspace{-5pt} \begin{array}{c}
      q( t + \delta \tau ) \\
      p( t + \delta \tau )
	\end{array} \hspace{-5pt} \right] = \left[ \hspace{-5pt} \begin{array}{cc}
	   1 - \frac{1}{2}\left( \omega \delta \tau \right)^2 & \omega \delta \tau \\
	  -\omega \delta \tau + \frac{1}{4}\left(\omega \delta \tau \right)^3 & 1 - \frac{1}{2}\left( \omega \delta \tau \right)^2  
	\end{array} \hspace{-5pt} \right] \left[ \hspace{-5pt} \begin{array}{c} q(t) \\ p(t) \end{array} \hspace{-5pt} \right] \label{e:SHOUpdate}
\end{equation}
where $q$ and $p$ are the coordinates and conjugate momenta, $t$ is MD time and $\omega$
is the angular frequency.

The eigenvalues of the matrix in (\ref{e:SHOUpdate}) are 
\begin{equation}
\lambda_{\pm} =  e^{\pm i \cos^{-1}\left(1 - \frac{1}{2}\left( \omega \delta \tau \right)^2 \right)} \label{e:EigVal}
\end{equation}
 
When $\omega \delta \tau < 2$ the inverse cosine in (\ref{e:EigVal})
is real, the eigenvalues are complex, and the phase space trajectories
describe ellipses. However, when $\omega \delta \tau > 2$, the inverse
cosine becomes imaginary, and one eigenvalue is greater than one. At
this point the phase space trajectories become hyperbolic, and the
energy change along a trajectory grows exponentially.

The hypothesis of \cite{Edwards} is that {\em the short distance modes
of an asymptotically free quantum field theory behave like a collection
of loosely coupled SHO modes}. It is then anticipated that the MD
integration of lattice QCD will go unstable when the highest frequency
mode goes unstable.

We note that in the SHO example $\omega$ plays the role of the MD
force $F$. Hence we surmise that for QCD the instability will occur
when the combination $F \delta \tau$ reaches a critical
value. Furthermore, we expect that the fermionic contribution to the
force will increase in some manner with the inverse quark mass. This
suggests that as the quark masses are decreased, the $F \delta \tau$
term will reach its critical value for smaller values of $\delta \tau$.
This indeed is exactly what is reported in \cite{Edwards} and \cite{Joo}.

\section{NUMERICAL RESULTS}\label{s:OurData}
Our numerical computations were carried out using 10 lattice
configurations from a UKQCD lattice computation \cite{UKQCD}. The lattices were
picked evenly spaced from a larger ensemble. The lattice volume was
$16^3\times32$ sites, and the physical parameters for the
ensemble were $\beta=5.2$ for the gauge coupling,
$\kappa=0.1355$ for the hopping parameter and $c_{\rm SW}=2.0171$ for
the $O(a)$--improvement coefficient \cite{SW,Jansen}. These parameters  correspond to a regime where the ratio of the pseudoscalar to vector masses is
$\frac{m_{\pi}}{m_{\rho}} \approx 0.6$ \cite{UKQCD,Joyce}.

We evolved the configurations for unit length
MD trajectories using a variety of step sizes at a 
variety of $\kappa$ values.  Along each trajectory
we computed $\delta H$ as well as the 2-norms and $\infty$
norms of the gauge and fermionic contributions to the MD force ($||F_{g}||_{2}$,$||F_{f}||_2$, $||F_{g}||_{\infty}$, and $||F_{f}||_{\infty}$) averaged over all the time steps along the trajectory.

\begin{figure}[ht]
\includegraphics[width=18pc]{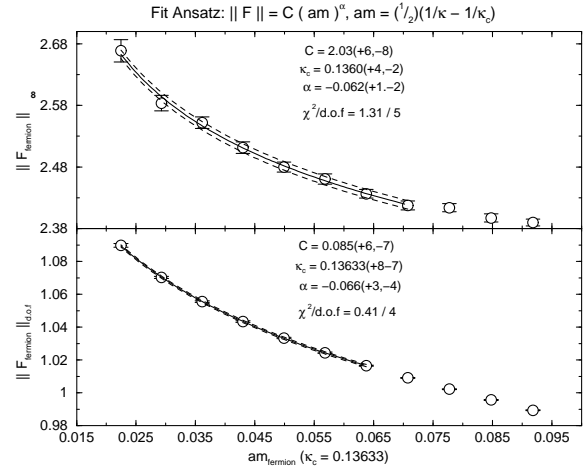}
\vspace{-1.3cm}
\caption{$||F_{f}||_2$ and $||F_{f}||_{\infty}$ vs $am$.}
\vspace{-0.7cm}
\label{f:Forces_am}
\end{figure}

In figure \ref{f:Forces_am} we show the results of fitting the fermionic
force norms (averaged over configurations) to the fit ansatz:
\begin{equation}
||F|| = C(am)^{\alpha}
\end{equation}
in order to check that the forces increase in some inverse manner
with the quark mass. To calculate $am$ we used the formula
\begin{equation}
am = \frac{1}{2}\left( \frac{1}{\kappa} - \frac{1}{\kappa_{\rm crit}} \right) \
\end{equation}
where $\kappa_{\rm crit}$ is the critical value of $\kappa$
corresponding to the massless limit of the pion. In our fits
we have left $\kappa_{\rm crit}$ as a free parameter, however to
determine $am$ for the axes of figure \ref{f:Forces_am} we used the value
$\kappa_{\rm crit} = 0.13633$, determined from spectroscopy
on the ensemble \cite{Joyce}.

One can see from figure \ref{f:Forces_am} that the values of $\alpha$ 
are negative for both the 2--norm and the $\infty$--norm indicating that
the forces do indeed grow inversely with $am$. The 
fits for $\kappa_{\rm crit}$ are consistent with the spectroscopic determinations indicating that the forces will diverge as $am \rightarrow 0$. 

%\begin{figure}[htb]
%\includegraphics[width=18pc]{Forces_vs_dtau.eps}
%\vspace{-1.3cm} 
%\caption{2-norms (bottom), $\infty$-norms (middle) and corresponding
%values of $\delta H$ (top) vs $\delta \tau$.}
%\vspace{-0.7cm}
%\label{f:Forces_dtau}
%\end{figure}

\begin{figure}[htb]
\includegraphics[width=18pc]{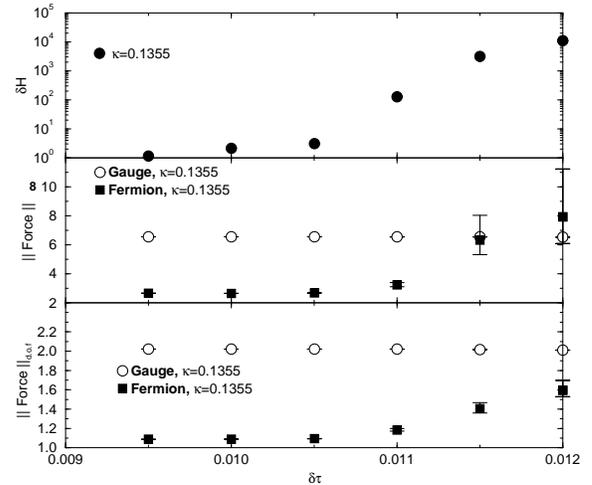}
\vspace{-1.3cm} 
\caption{2-norms (bottom), $\infty$-norms (middle) and corresponding
values of $\delta H$ (top) vs $\delta \tau$.}
\vspace{-0.7cm}
\label{f:Forces_dtau}
\end{figure}

In figure \ref{f:Forces_dtau}, we plot the variation of the force norms
and $\delta H$ against the step size $\delta \tau$ at a fixed 
value of $\kappa = 0.1355$ for a single configuration. It can be seen that 
at a value of about $\delta \tau = 0.0105$, the fermionic forces start increasing and that $\delta H$ increases exponentially indicating the onset 
of the instability. The gauge contributions to the force norms
seem to show no change in behaviour. 

The $\infty$--norms grow at a larger rate than the 2--norm. As
the $\infty$--norm is the largest absolute value contribution to the
fermionic force, it plays a similar role to the highest frequency mode
of a corresponding set of loosely coupled SHOs whereas the 2--norm of
the force can be likened to the behaviour of the corresponding bulk
modes. Hence this result is consistent with the earlier hypothesis,
that the highest frequency modes drive the system unstable.

\begin{figure}[tb]
\includegraphics[width=18pc]{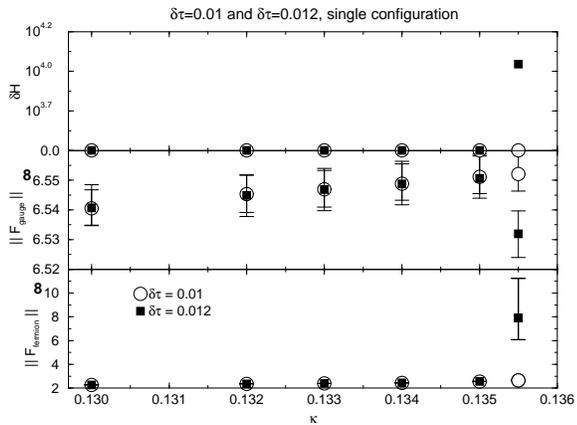}
\vspace{-1.3cm}
\caption{$||F_{f}||_{\infty}$ (bottom), $||F_{g}||_{\infty}$ (middle), $\delta H$ (top) vs $\kappa$, for two values of the step size: $\delta \tau = 0.01$ (circles) and $\delta \tau = 0.012$ (squares).}
\vspace{-0.5cm}
\label{f:Forces_kappa}
\end{figure}

In figure \ref{f:Forces_kappa} we show the behaviour of
$||F_{f}||_{\infty}$, $||F_{g}||_{\infty}$ and $\delta H$ against the hopping
parameter $\kappa$ for $\delta \tau = 0.01$
and $\delta \tau = 0.012$. When $\delta \tau = 0.01$ the system is
completely stable for all values of $\kappa$ (for this single configuration); however, for $\delta
\tau = 0.012$ the fermionic force norm rises sharply
for $\kappa = 0.1355$ accompanied by a large increase in $\delta H$
signalling the onset of the instability. This again is
consistent with our earlier hypothesis that instabilities set in when
the $F \delta\tau$ term reaches a critical value.

\section{CONCLUSIONS AND DISCUSSION}\label{s:Conclusions}
We have exhibited the onset of an instability in the leapfrog MD integration
scheme, used in HMC simulations and inexact simulation
algorithms. The instability is manifest in the leapfrog scheme for a simple
SHO system. We have also presented numerical evidence demonstrating that the 
instability is present also in lattice QCD systems. We have shown that our
data is consistent with the hypothesis that the instability  occurs when
the $F \delta \tau$ term in the leapfrog equations reaches some critical value, and that the fermionic contribution to the force increases
with decreasing quark mass. We have shown that as quark masses become lighter
the instability sets in at smaller values of $\delta \tau$. Furthermore, 
for simulations with light dynamical quarks the instability sets in at 
values of $\delta \tau$ that are small enough to be relevant to large scale numerical simulations.

We anticipate that the major stumbling block for simulations with lighter 
quark masses will not come from ``exceptional'' configurations, where the 
value of $\kappa$ may become supercritical for some configurations, but 
rather from the onset of the instability which we expect to set in before $\kappa$ can become supercritical.

The instability problem is exponentially bad, but can be controlled easily by reducing $\delta \tau$. We are concerned for simulations
with inexact algorithms, where the instability may be hard to detect as there
is no $\delta H$ to monitor, and full finite step size extrapolations are 
seldom made. 

Finally we wish to mention that neither the use of higher order integration schemes nor the use of double precision arithmetic are expected to alleviate the instability problem \cite{Joo}.

\section{ACKNOWLEDGEMENTS}
We acknowledge support from PPARC under grant number GR/L22744.


\begin{thebibliography}{9}
\bibitem{Edwards} R.~G.~Edwards, I.~Horv\'ath, A.~D.~Kennedy, {\em Nucl. Phys.\ }{\bf B484} (1997), 375-402
\bibitem{Joo} B.~Jo\'o {\em et. al.}, {\em Phys. Rev.\ }{\bf D62}(2000), 114501
\bibitem{UKQCD} C.~R.~Allton {\em et. al.}, {\tt hep-lat/0107021}
\bibitem{SW} B.~Sheikholeslami, R.~Wohlert, {\em Nucl. Phys.\ }{\bf B259}, (1985)572.	
\bibitem{Jansen} K.~Jansen, R.~Sommer, {\em  Nucl. Phys. B.(Proc. Suppl.)\ }{\bf 63A-C}(1998) 853-855
\bibitem{Joyce} J.~Garden, PhD Thesis, Edinburgh University, 2000
\end{thebibliography}
\end{document}